\documentclass[11pt,tightenlines,notitlepage,longbibliography,aps,prd]{revtex4-1}

\usepackage[T1]{fontenc}
\usepackage[utf8]{inputenc}
\usepackage{amssymb}
\usepackage[english]{babel}
\usepackage[dotinlabels]{titletoc}
\usepackage{lmodern}
\usepackage{amsmath}
\usepackage[font=small,labelfont=bf]{caption}
\usepackage{graphicx}
\usepackage{cancel}
\usepackage{bm}
\usepackage{courier}
\usepackage{hyperref}
\usepackage{float}

\newcommand{\der}[2]{\frac{\text{d} #1}{\text{d} #2}}

\begin{document}

\title{Towards a Hartle-Hawking state for loop quantum gravity}

\author{Satya Dhandhukiya}
\email{satya.dhandhukiya@gravity.fau.de}
\author{Hanno Sahlmann}
\email{hanno.sahlmann@gravity.fau.de}

\affiliation{\textit{Institute for Quantum Gravity, Department of Physics, Friedrich-Alexander-Universit\"at Erlangen-N\"{u}rnberg (FAU), Erlangen (Germany)}}

\begin{abstract}
The Hartle-Hawking state is a proposal for a preferred initial state for quantum gravity, based on a path integral over all compact Euclidean four-geometries which have a given three-geometry as a boundary. The wave function constructed this way satisfies the (Lorentzian) Hamiltonian constraint of general relativity in ADM variables in a formal sense. In this article we mimic this procedure of constructing an initial state in terms of Ashtekar-Barbero variables, and observe that the wave function thus constructed does not satisfy the Lorentzian Hamiltonian constraint even in a formal sense. We also investigate this issue for the relativistic particle. We finally suggest a modification of the proposal that does satisfy the constraint at least in a formal sense and start to consider its implications in quantum cosmology. We find that for certain variables, and in the saddle point approximation, the state is very similar to the Ashtekar-Lewandowski state of loop quantum gravity.  
\end{abstract}	

\maketitle

\section{Introduction}
An important motivation in the search for a quantum theory of gravity is the resolution of the singularities present in generic solutions to classical general relativity. Of particular interest in this context is the big bang singularity, which is intimately connected to the conditions at the beginning of the Universe. Consequently the question of singularity resolution is related to that of the initial conditions in quantum gravity. 

In \cite{Hartle:1983ai}, Hartle and Hawking (HH) proposed an extremely elegant definition for an initial state for canonical quantum gravity, 
\begin{equation}\label{eq9}
\Psi^{\text{HH}}_0[q_{ab}] := \int_{D(q_{ab})}  \mathcal{D}[g]\: e^{-S_E[g]}.
\end{equation}
This proposal is within the ADM approach to quantum gravity, i.e., states are wave functions of the spatial metric $q_{ab}$. The integration domain $D(q_{ab})$  of the path integral contains all smooth Euclidean 4-metrics on a compact manifold with boundary, such that the metric on the boundary is $q_{ab}$. The weighting factor is given by the Euclidean action $S_E(g)$, and the over-counting due to diffeomorphism-related metrics is assumed to be taken care of by the path integral measure $\mathcal{D}[g]$. This state satisfies the constraints of canonical quantum gravity, in particular the Wheeler-de Witt equation, in a formal sense. While it is hard to go beyond formal considerations in the full theory due to the difficulties involved with defining the path integral, the no boundary proposal is amenable to analytic and numerical treatment in the case of quantum cosmology and has been studied there in great detail, see \cite{Gibbons:1994cg} and the references therein, for example. Some more recent developments are in \cite{Hartle:2007gi,Hartle:2008ng,Hartle:2010vi,Hwang:2011mp,Hwang:2012zj,Hwang:2012bd,Hwang:2014vba}. 

Loop quantum gravity (LQG) is a canonical approach to the quantization of gravity, making use of an extension of the ADM phase space that is embedded in the phase space of SU(2) Yang-Mills theory \cite{Ashtekar:1986yd,Barbero:1994ap}. While the issue of singularity resolution is not settled in full generality in LQG, there are many indications that the quantum theory does resolve the singularities of general relativity (GR). For example, there are indications that the big bang singularity is replaced by a bounce from a minimal scale factor. 

This still leaves the question of what the appropriate initial state is in LQG. It has started to become relevant in loop quantum cosmology \cite{Bojowald:2001xe,Ashtekar:2003hd}. It has been shown that a large class of initial states is compatible with observations, \cite{Agullo:2012sh}, but that details of the initial state may nevertheless be probed by future observations \cite{Agullo:2013ai}. This is our primary motivation to come back to the HH proposal and study its applicability in LQG. 

The definition of the Hartle-Hawking state is intimately tied to the ADM formulation of gravity. Since different variables are used in loop quantum gravity, the translation of the proposal is non-trivial. We investigate various straightforward possibilities, but find that none of the resulting states satisfies the Hamiltonian constraint. We finally partially specify a state that does satisfy the constraint at least formally. The proposed state differs in important respects from the original proposal by Hartle and Hawking, so that its physical and mathematical viability has to be considered from scratch. We start work in this direction by considering cosmological models. Since the difficulties in translating the HH proposal are somewhat surprising at first sight, we also study the situation in a toy model: The free relativistic particle. 

We should point out that path integrals for quantum cosmology in the LQG context have been discussed in detail in \cite{Ashtekar:2010gz}. There the main object of study is the \emph{extraction kernel}, which in ADM variables would read 
\begin{equation}
\label{eq_extract}
\Psi[q^{(1)}_{ab},q^{(2)}_{ab}] := \int_{D(q^{(1)}_{ab},q^{(2)}_{ab})}  \mathcal{D}[g]\: e^{iS [g]}.
\end{equation}
It can be considered the analogue to the propagator in a theory in which the canonical Hamiltonian is a constraint.  In this case the underlying manifold has a boundary consisting of two disjoint pieces, and the integration domain $D(q^{(1)}_{ab},q^{(2)}_{ab})$ is defined by prescribing the geometries on those pieces. In \cite{Ashtekar:2010gz}, the projector on the kernel of the quantum constraint is investigated in detail in the cosmological setting. It is shown that it has a Feynman-like path integral representation, with an action that carries quantum corrections. The present work is much more formal than \cite{Ashtekar:2010gz} in that we do not start from the mathematically rigorous Hilbert space underlying LQG, and consequently, we will work with the classical action, and not consider quantum corrections. Moreover, extrapolation from the properties of the usual propagator for particles in quantum theory suggests that the extraction kernel is not a natural ground state in any sense. 

We should finally point out that an interesting proposal that carries some formal similarities to the Hartle-Hawking state has been put forward in the spin foam formulation of loop quantum gravity \cite{Bianchi:2010zs}. The precise relation to the states considered in the present work is not clear, however. 

The article is structured as follows: In section \ref{sec_adm} we review the work of Hartle-Hawking and give a precise meaning of what their proposal entails in ADM variables. In section \ref{sec_ab}, we try to translate this proposal in the context of loop quantum gravity and see that all straightforward proposals fail. In section \ref{sec_new}, we make a  new proposal for an initial state that does satisfy the constraint equations of loop quantum gravity. Finally, in section \ref{sec_lqc}, we apply this new state to (spatially flat) loop quantum cosmological models and perform a saddle point approximation. We finish this article with a summary and an outlook on future research directions. In an appendix, we consider the case of a relativistic particle as a toy model in order to see the difficulties that arise in proposing an initial state.

\textit{Conventions}: In what follows, we will consider a spacetime manifold $M$ with boundary $\partial M$. The metric convention of $(-\,+\,+\,+)$ for Lorentzian signature and $(+\,+\,+\,+)$ for Euclidean signature is used. To write formulas for both signatures in a unified way, we will make use of the variable $s$ taking the value $-1$ in the Lorentzian and $+1$ in the Euclidean case. 
Greek indices $\mu,\nu,\rho,\ldots=0,1,2,3$ are used for components of spacetime tensors, latin indices $a,b,c,\ldots=1,2,3$ for spatial tensors. $i,j,k,\ldots =1,2,3$ label components in the adjoint representation of SU(2) and $I,J,K,\ldots=1,2$ in the defining representation of SU(2). 
\section{Hartle-Hawking state for ADM variables}
\label{sec_adm}
We use the Einstein-Hilbert action (setting $16\pi G=1$) in the form 
\begin{equation}\label{eq11}
S = \int_{M}d^{4}x\,\sqrt{|\det(g)|}\,R^{(4)} + \frac{1}{2}\int_{\partial M} d^{3}x\, \sqrt{|\det(q)|} K. 
\end{equation}
We have prepared for the fact that we will consider manifold $M$ with boundary, by including the Gibbons-Hawking-York boundary term.  Here $R^{(4)}$ is the 4-Ricci scalar on spacetime manifold $M$, $q_{ab}$ is the metric induced on the boundary and $K$ is the trace of the extrinsic curvature on the boundary. 

The proposal \eqref{eq9} by Hartle and Hawking \cite{Hartle:1983ai} for the initial state of geometry is based on the ADM formalism \cite{Arnowitt:1959ah}. It involves a foliation of spacetime into spatial slices diffeomorphic to a 3d manifold $\sigma$. We assume that the boundary is spacelike, and that the foliation is adapted to the boundary in the sense that the boundary is given by one of the slices. In adapted coordinates the metric takes the form 
\begin{equation}\label{eq10}
ds^2 = (sN^2 + q_{ab} N^a N^b)dt^2 + 2q_{ab} N^b dt^a dx^b + q_{ab} dx^a dx^b.
\end{equation}
$N$ is the lapse function, $N^a$ is the shift, $q_{ab}(x)$ is the 3-metric on $\sigma$ and $s$ is the signature of spacetime, with $s=-1$ for Lorentzian and $s=1$ for Euclidean signature.  
With this split, the action can be written in terms of the ADM variables as \cite{Arnowitt:1962hi, Thiemann:2007zz,poisson2004relativist}
\begin{equation}\label{eq12}
S=\int^{t_0}dt\int_{\sigma}d^{3}x\,\sqrt{\det(q)}\,|N|\,\left[R^{(3)}-s\,\left(K_{ab}K^{ab}-({K^a}_a)^2\right)\right],
\end{equation}
where $R^{(3)}$ is the 3-Ricci scalar on $\sigma$, $K_{ab}$ is the extrinsic curvature of $\sigma$ in spacetime and ${K^a}_a$ is the trace of the extrinsic curvature. $t_0$ is the time coordinate of the slice $\partial M=\sigma$. 

We will also use the covariant derivative $D$ associated to the spatial metric $q_{ab}$. 
The boundary term in \eqref{eq11} is chosen such that there is no boundary contribution in \eqref{eq12}. 

Using $q_{ab}, N, N^a$ as variables the conjugate momenta are:
\begin{align}\label{eq13}
P^{ab}(t,x)&:= \frac{\delta S}{\delta {\dot{q}_{ab}}} = -s\,\frac{|N|}{N}\,\sqrt{\det(q)}\,\left[(K^{ab}-q^{ab}({K^c}_c))\right],\\
\Pi(t,x)&:= \frac{\delta S}{\delta \dot{N}}=0,\qquad
\Pi_a(t,x):= \frac{\delta S}{\delta \dot{N}^a}=0
\end{align}
which shows that $N,N^a$ play the role of Lagrange multipliers. Rewriting $S$ in terms of  $q,P, N, N^a$ one finds 
\begin{equation}\label{eq18}
S = \int_{\mathbb{R}}dt\int_{\sigma}d^{3}x\,\left\{\dot{q}_{ab}P^{ab}\,-\,\left[N^{a}H_a+|N|H\right]\right\}
\end{equation}
where 
\begin{align}\label{eq16}
H_a&=-2q_{ac}D_{b}P^{bc},\\
H&=-\left(\frac{s}{\sqrt{\det(q)}}\left[q_{ac}q_{bd}-\frac{1}{2}q_{ab}q_{cd}\right]P^{ab}P^{cd}\,+\,\sqrt{\det(q)}\,R^{(3)}\right),
\end{align}
are the \textit{(spatial) diffeomorphism constraint} and \textit{Hamiltonian constraint}, respectively. In the following we will have to deal with both signatures, so we will also introduce the notation 
\begin{equation}
H_L=H\rvert_{s=-1}, \qquad H_E=H\rvert_{s=+1}. 
\end{equation}
A formal canonical quantization proceeds by stipulating a Hilbert space 
\begin{equation}
\mathcal{H}_{\text{ADM}}=L^2(\mathcal{Q}, d\mu) 
\end{equation}
with $\mathcal{Q}$ a space of 3-metrics and $d\mu$ a uniform measure on this space. Wave functions are thus functionals of 3-metrics, and the operators  
\begin{equation}\label{eq19}
\widehat{Q}_{ab}(x)\Psi[q]=q_{ab}(x)\Psi[q], \qquad \widehat{P}^{ab}(x)\Psi[q]=\frac{1}{i}\,\frac{\delta}{\delta {q}_{ab}(x)}\Psi[q]
\end{equation}
are assumed to be self-adjoint and fulfil the canonical commutation relations. Here, and in this article, we set $\hbar =1$. 

The observation of Hartle-Hawking is that the Lorentzian Hamiltonian constraint acting on the state \eqref{eq9} vanishes in the formal quantization given above, i.e., 
\begin{equation}
\label{eq_physical}
\widehat{H}_L \Psi^{\text{HH}}_0[{q}_{ab}]=0
\end{equation}
where $\widehat{H}_L $ is obtained from the classical expression by inserting the operators \eqref{eq19} in a suitable order.
The argument to show \eqref{eq_physical} proceeds in two steps. First, one can express $g_{ab}$ in terms of the ADM variables $q_{ab}$, $N^a$, $N$, and use the action in ADM form \eqref{eq18}, 
\begin{equation}
 \Psi^{\text{HH}}_0[{q}_{ab}]=\int_{D(q_{ab})}  \mathcal{D}q'_{ab}\,\mathcal{D}N^a\,\mathcal{D}N\: e^{-S_E[q'_{ab},N^a,N]}
\end{equation}
where $D(q_{ab})$ is now a domain in the space of ADM variables that enforces the boundary conditions as before. Upon the natural assumption that the measure $DN$ is uniform under arbitrary translations in the space of lapse functions, 
\begin{align}\label{eq_a}
\int_{D(q_{ab})}   \mathcal{D}q'_{ab}\,\mathcal{D}N^a\,\mathcal{D}N\: H_E(f)\:e^{-S_E[q'_{ab},N^a,N]}&=\left. \der{}{\epsilon}\right\rvert_{\epsilon=0} \int_{D(q_{ab})}   \mathcal{D}q'_{ab}\,\mathcal{D}N^a\,\mathcal{D}N\:  e^{-S_E[q'_{ab},N^a,N+\epsilon f]}\\
&= \left. \der{}{\epsilon}\right\rvert_{\epsilon=0} \int_{D(q_{ab})}   \mathcal{D}q'_{ab}\,\mathcal{D}N^a\,\mathcal{D}N\:  e^{-S_E[q'_{ab},N^a,N]}\\
&=0, \label{eq_b}
\end{align}
where $f$ is an arbitrary function on the boundary $\partial M$ and we have used the notation 
\begin{equation}
H_E(f)=\int_{\partial M} H_E(x)f(x).
\end{equation}
With the same argument, one can also show that insertion of the classical diffeomorphism constraint under the above path integral will lead to a vanishing integral. 

The second step consists in showing that the quantum \emph{Lorentzian} Hamiltonian constraint turns into the classical \emph{Euclidean} Hamiltonian constraint under the path integral, 
\begin{equation}
\widehat{H}_L(x)\,\Psi^{\text{HH}}_0[{q}_{ab}]=\int_{{D}(q_{ab})}   \mathcal{D}q'_{ab}\,\mathcal{D}N^a\,\mathcal{D}N\: H_E(x)\:e^{-S_E[q'_{ab},N^a,N]}. 
\end{equation}
Let us first consider the action of $\widehat{P}$ on $\Psi^{\text{HH}}_0$. The most important thing to note is that the functional derivative in $P$ just concerns the boundary value ${q}_{ab}=q'_{ab}(t_0)$ of the histories $q'_{ab}(t)$ that are integrated over in the path integral. Taking the derivative thus requires some care. We will do it by first taking a standard variation of the action $S$ with respect to $q'_{ab}$. For this variation, we treat $q'$ and $P$ as independent variables. 
In a second step we then take a limit in which the variation becomes restricted to the boundary. We have 
\begin{equation}\label{eq_mess}
\left[\int d^4x\; h_{ab}(x)\,\frac{1}{i}\,\frac{\delta}{\delta q_{ab}(x)} \right]\; S[P,q']
= \int^{t_0}dt\int_{\sigma} d^3x\;  \left[-i \dot{h}_{ab}P^{ab} -\,i h_{ab}\frac{\delta}{\delta q_{ab}} (NH+ N^cH_c)\right].
\end{equation}
We have not written out the functional deriviative in the second term explicitly because we will see momentarily that it does not contribute. We take a suitable limit 
\begin{equation}\label{eq_lim}
h_{ab}(t, x)\longrightarrow f_{ab}(x)
\begin{cases} 1 & \text{ if } t=t_0\\
0 & \text{ otherwise }
\end{cases}
\end{equation}
which concentrates the variation on the boundary $t=t_0$. Since $H$ and $H_a$ do not contain time derivatives of $q$, the terms resulting from the functional derivative will go to zero in the above limit. They are bounded functions with support concentrated more and more on the boundary. Contrary to this, the first term on the right hand side of \eqref{eq_mess} hides a boundary term which does not vanish in the limit: 
\begin{equation}
\int^{t_0}dt\int_{\sigma} d^3x\; \dot{h}_{ab}P^{ab}=[h_{ab}P^{ab}]^{t_0}-\int^{t_0}dt\int_{\sigma} d^3x\; {h}_{ab}\dot{P}^{ab}.
\end{equation}
The second term vanishes in the limit \eqref{eq_lim}, but the first term does not. That is 
\begin{equation}
\left[\int d^4x\; h_{ab}(x) \frac{\delta}{\delta q_{ab}(x)} \right]\; S[P,q'] \longrightarrow P^{ab}(t_0)h_{ab}(t_0) .
\end{equation}
Therefore we obtain the simple result 
\begin{equation}
\widehat{P}^{ab}(x)\,\Psi^{\text{HH}}_0[{q}_{ab}]=\int_{{D}(q_{ab})}   \mathcal{D}q'_{ab}\,\mathcal{D}N^a\,\mathcal{D}N\:  iP^{ab}(x)\:e^{-S_E[q'_{ab},N^a,N]}. 
\end{equation}
Now one can show \eqref{eq_physical}:
\begin{align}
\widehat{H}_L\,\Psi^{\text{HH}}_0[{q}_{ab}]
&=\left(\frac{1}{\sqrt{\det(\widehat{q})}}\left[\widehat{q}_{ac}\widehat{q}_{bd}-\frac{1}{2}\widehat{q}_{ab}\widehat{q}_{cd}\right]\widehat{P}^{ab}\widehat{P}^{cd}\,-\,\sqrt{\det(\widehat{q})}\,\widehat{R}^{(3)}\right)\:\int_{D(q)} \mathcal{D}[g]\: e^{-S_E[g]}\\
&=\int_{D(q)}\mathcal{D}[g]\left[-\left(\frac{1}{\sqrt{\det({q})}}\left[{q}_{ac}{q}_{bd}-\frac{1}{2}{q}_{ab}{q}_{cd}\right]{P}^{ab}{P}^{cd}\,+\,\sqrt{\det({q})}\,{R}^{(3)}\right)\right]\: e^{-S_E[g]}\label{eq22}\\
&=\int_{D(q)}\mathcal{D}[g]\:H_E\:e^{-S_E[g]}.\label{eq23}
\end{align}
Note that there is a combination of sign factors that lead to this result: On the one hand, $\widehat{P}\widehat{P}$ goes to $-PP$ under the path integral. On the other hand, the $PP$ term in the Hamiltonian constraint comes with a factor $s$ and thus changes sign. We will see momentarily that such a combination of signs does not take place when using connection variables for gravity. 

A calculation similar to the one above shows that 
\begin{equation}
\widehat{H}_{a}\,\Psi^\text{HH}_0[{q}_{ab}]=0.
\end{equation}
In this case, no fortuitous combinations of signs is necessary to reach the result. 
\section{Hartle-Hawking state for Ashtekar-Barbero variables}
\label{sec_ab}
Given the construction outlined in the last section, it is a natural question to ask whether an analogous state can be defined for loop quantum gravity, at least at the formal level. We will see that this is a non-trivial question, and that the most natural way to generalize the construction to the variables used in LQG results in a state that does not satisfy the constraints in a formal sense. 

LQG starts from Ashtekar variables  \cite{Ashtekar:1986yd} in their real form \cite{Barbero:1994ap}. These embed the gravity phase space into the phase space of SU(2) Yang-Mills theory. We use the conventions and results of \cite{Thiemann:2007zz}. The set of variables are the electric field $E^a_i$ and the \textit{Ashtekar-Barbero connection} $A^j_b$, which satisfy the canonical commutation relations, 
\begin{equation}\label{eq24}
\{E^a_i(x)\,,\,A^j_b(y)\} = \frac{1}{2}\:\delta^a_b\,\delta^j_i\,\delta^3(x-y).
\end{equation}
The connection $A^j_b$ is defined as $A^j_b:=\Gamma^j_b\,- s\,\gamma K^j_b$, where $\Gamma$ is the \textit{spin connection}, $\gamma$ is called the \textit{Immirzi parameter} which can take any non-zero real value, $s$ is the signature of the spacetime manifold $M$ as above, and $K$ is the extrinsic curvature.

These variables \cite{Thiemann:2007zz,Ashtekar:2004eh} can be derived from the Holst action \cite{Thiemann:2007zz,Holst:1995pc,Ashtekar:2004eh}. In its canonical form it reads
\begin{equation}\label{eq29}
S = \int_{\mathbb{R}}dt\int_{\sigma}d^{3}x\,\{2\dot{A}^i_a E^a_i\,-\,[{\Lambda^j}{G_j}\,+ {N^a}H_a\,+NH]\}
\end{equation}\\
with the three constraints:
\begin{align}
G_j &= \frac{D_a\,E^a_j}{\gamma},\label{eq30}\\
H_a &=- \frac{s}{\gamma} F^j_{ab}E^b_j,\label{eq31}\\
H &= [F^j_{ab}\,-\,({\gamma}^2 - s)\epsilon_{jmn}\,K_a^m K_b^n]\frac{\epsilon_{jkl}E^a_k E^b_l}{\sqrt{\det(q)}}.\label{eq32}
\end{align}
The curvature of the connection $A^j_b$ is $F^j_{ab}$, and it is given by $F^j_{ab}=\partial_a A^j_b\,-\partial_b A^j_a\,+\epsilon^j_{mn} A^m_a A^n_b$. 

To obtain \eqref{eq29} from the covariant form of the Holst action in the presence of boundaries, suitable boundary terms (generalizing the Gibbons-Hawking-York boundary term of \eqref{eq11}) have to be added. The question of what the appropriate boundary terms are for the Holst action has only been addressed recently, see for example \cite{Corichi:2016zac}. Here we will just \emph{assume} that, by the addition of suitable boundary terms, the action has been brought into the form \eqref{eq29}. 

Note that in addition to the diffeomorphism constraint $H_a$ and the Hamiltonian constraint $H$, one has an additional constraint $G_j$ called the \textit{Gauss constraint}. This constraint arises as an extra feature of the theory due to the internal SU(2) gauge group which generates gauge transformations on the phase space.
In finding the expression \eqref{eq29} for the Holst action analogous to \eqref{eq18}, we have freely added boundary terms to the original action, to be able to carry out the required partial integrations.  

We want to construct a ground state function which mimics the properties of the Hartle-Hawking state. But before doing so, we are faced with a choice of the variable that needs to be considered in defining the initial state: We could fix the connection $A$ or the electric field $E$ on the boundary $\partial M$. Hartle's and Hawking's construction could be generalized in two ways: By focusing on the fact that $q_{ab}$ plays the role of a \emph{canonical position} variable or, by focusing on the fact that $q_{ab}$ on the boundary defines its \emph{intrinsic geometry}. The former would lead to a wave function of $A$, the latter to one of $E$. In fact, in the former case where the state is a function of the connection $A$, a simple canonical transformation would make $E$ the canonical position variable. For a linear system such as a harmonic oscillator, these two choices can be related by a Fourier transform, but this is an artefact of the simple nature of the system, and does no longer hold for gravity. Finally, there could be the possibility of using the original state in a construction of a ground state for LQG. 

Therefore, the three wave functions, respectively, would read
\begin{equation}
\Psi^{\text{LQG}}_1[A], \qquad \Psi^{\text{LQG}}_2[E], \qquad\text{and}\qquad \Psi^{\text{LQG}}_3[E]:=\Psi^{\text{HH}}_0[q(E)],
\end{equation}
where the superscript LQG refers to the fact that we are now dealing with states within loop quantum gravity.

We will now deal with all of these possibilities one by one. We will see that for each of these possibilities, the resulting state does not satisfy the (Lorentzian) Hamiltonian constraint. Since this difficulty is somewhat surprising, we also study the situation in a toy model, the free relativistic particle, in an appendix. The result is the same.  
\subsection{Wave function of $A$: $\Psi^{\text{LQG}}_1$}
\label{se_psiofa}
Following the Hamiltonian formulation and considering the fact that the role these two variables play is analogous to the position and momentum variables, we start out by defining the state in terms of the connection. 
A formal quantization of \eqref{eq24} can be obtained by stipulating a Hilbert space 
\begin{equation}
\label{eq_hilb}
\mathcal{H}=L^2(\mathcal{A}, d\mu) 
\end{equation}
with $\mathcal{A}$ a space of connections on $\sigma$ and $d\mu$ a uniform measure on this space. Wave functions are thus functionals of $A$, and the operators  
\begin{equation}
\widehat{A}_{a}^i(x)\Psi[A]=A_{a}^i(x)\Psi[A], \qquad \widehat{E}^{a}_i(x)\Psi[A]=\frac{1}{2i}\,\frac{\delta}{\delta {A}_{a}^i(x)}\Psi[A]
\end{equation}
are assumed to be self-adjoint and fulfill the canonical commutation relations
\begin{equation}
\label{eq_comm}
[E^a_i(x)\,,\,A^j_b(y)] = \frac{i}{2}\:\delta_b^a\,\delta_i^j\,\delta^3(x,y).
\end{equation}
LQG is based on a mathematically rigorous version of the Hilbert space \eqref{eq_hilb} and the canonical commutation relations \eqref{eq_comm} \cite{Ashtekar:1994mh,Ashtekar:1994wa,Ashtekar:1995zh}. In the present exploratory work, we will stay on the formal level. 

To obtain a state as a function of the connection, the integration in the definition of the state \eqref{eq9} must be over all possible histories of connections $A'(t,x)$ that have a common boundary value $A(x)=A'(t_0,x)$. These histories form a set $D(A)$. Here we assume, as in the ADM case, that the only boundary of the manifold is $\sigma$, and that $A$ and $\dot{A}$ are smooth (``no boundary condition''). We will not explore what this actually means for the form on $A$, as we will see that serious problems already arise at a more elementary level. 
Hence we tentatively define the quantum state as
\begin{equation}\label{eq33}
\Psi^{\text{LQG}}_1[A] := \int_{D(A)}  \mathcal{D}A'\,\mathcal{D}N^a\,\mathcal{D}N\: e^{-S_E[A',N^a,N]}.
\end{equation}
With the same arguments as in the ADM case, one can show that
\begin{equation}
\label{eq_delta1}
\int_{D(A)}   \mathcal{D}A'\,\mathcal{D}N^a\,\mathcal{D}N\: H(f)\:e^{-S_E[A',N^a,N]}=0.
\end{equation}
Also, with the same arguments as in the ADM case, we can assume 
\begin{equation}
\widehat{E}_k^a(x)\,\Psi^{\text{LQG}}_1[A] =\int_{D(A)} \mathcal{D}A'\,\mathcal{D}N^a\,\mathcal{D}N\: iE^a_k(x)\, e^{-S_E[A',N^a,N]}.
\end{equation}
To find the action of the Hamiltonian constraint, note first that the extrinsic curvature in terms of the spin connection as:
\begin{equation}\label{eq36}
K^i_a=\frac{1}{s\gamma}(\Gamma^i_a\,-\,A^i_a).
\end{equation}
$\Gamma^k_a$ is a complicated function of $E^k_a$, but this function is rational and homogeneous of degree 0, i.e., it is invariant under scaling of $E^k_a$ \cite{Thiemann:2007zz}. Therefore
\begin{equation}
\widehat{\Gamma}_a^k(x)\,\Psi^{\text{LQG}}_1[A]=\int_{D(A)}   \mathcal{D}A'\,\mathcal{D}N^a\,\mathcal{D}N\: \Gamma_a^k(x)\, e^{-S_E[A',N^a,N]}.
\end{equation}
With these results, we can calculate the action of the Lorentzian Hamiltonian constraint on $\Psi^{\text{LQG}}_1$: 
\begin{align}
\widehat{H}_L\,\Psi^{\text{LQG}}_1[A]&=\left[\left[\widehat{F}^j_{ab}\,-\,({\gamma}^2 +1)\epsilon_{jmn}\,\widehat{K}_a^m \widehat{K}_b^n\right]\frac{\epsilon_{jkl}\widehat{E}^a_k \widehat{E}^b_l}{\sqrt{\det(\widehat{q})}}\right]\:\int_{D(A)}   \mathcal{D}A'\,\mathcal{D}N^a\,\mathcal{D}N\: e^{-S_E[A',N^a,N]}\label{eq37}\\
&=\int_{D(A)}\mathcal{D}A'\,\mathcal{D}N^a\,\mathcal{D}N\; \left[\left[-F^j_{ab}\,+\,({\gamma}^2 +1)\epsilon_{jmn}\,K_a^m K_b^n\right]\,\sqrt{i}\,\frac{\epsilon_{jkl}E^a_k E^b_l}{\sqrt{\det(q)}}\right]\; e^{-S_E[A',N^a,N]}\\
&=\int_{D(A)}\mathcal{D}A'\,\mathcal{D}N^a\,\mathcal{D}N\;\sqrt{i}\left[\frac{\epsilon_{jkl}E^a_k E^b_l}{\sqrt{\det(q)}}\left[2\epsilon_{jmn} K^m_a K^n_b\right]\right]\:e^{-S_E[A',N^a,N]}.
\end{align}
In the last step, \eqref{eq_delta1} has been used to eliminate a multiple of $H_E$ under the path integral. Unlike in the ADM case, a term remains under the integral. This term does not seem to be a simple function of the constraints. In particular  the form of the term is very different from the expressions for the Gauss and diffeomorphism constraints in \eqref{eq30} and \eqref{eq31}, respectively, and we have found no argument why it should have a vanishing integral. Therefore we conclude that $\Psi^{\text{LQG}}_1$ defined in this way \textit{does not} solve the Hamiltonian constraint. An interesting check of the formal calculations is given by applying the same reasoning to the extraction amplitude \eqref{eq_extract} in terms of Ashtekar-Barbero variables
\begin{equation}
\Psi^{\text{amplitude}}_1[A] := \int_{D(A)} \mathcal{D}A'\: e^{iS[A']}.
\end{equation}
Here one finds that applying $\widehat{A}$, $\widehat{E}$ to the state results in 
\begin{equation}
\widehat{A}(x)\leadsto   A(x), \qquad  \widehat{E}(x) \leadsto E(x), \qquad x \in \sigma 
\end{equation}
where it is understood that the operator on the left of the arrow results in the function on the right, inserted under the path integral. It is then obvious that 
\begin{equation}
\widehat{H}\,\Psi^{\text{amplitude}}_1[A]=0, \qquad \widehat{H}_a\, \Psi^{\text{amplitude}}_1[A]=0
\end{equation}
holds as it should -- both, for the Euclidean and for the Lorentzian case. 
\subsection{Wave function of $E$: $\Psi^{\text{LQG}}_2$}
We may also choose to define the ground state wave function as a functional of the field $E$. In this case, to make the analogy with the construction by Hartle and Hawking as strong as possible, we can regard $E$ as the position variable, i.e., work with the canonical pair $(E,-A)$. Then the appropriate formal quantization is given by
\begin{equation}\label{eq_hilb_alt}
\mathcal{H}=L^2(\mathcal{E}, d\mu) 
\end{equation}
with $\mathcal{E}$ a space of su(2) electric fields on $\sigma$ and $d\mu$ a uniform measure on this space. Wave functions are  functionals of E, and the operators  
\begin{equation}
\label{eq_ops2}
\widehat{E}^{a}_i(x)\Psi[E]=E^{a}_i(x)\Psi[E], \qquad \widehat{A}_{a}^k(x)\Psi[E]=\frac{i}{2}\,\frac{\delta}{\delta {E}^{a}_k(x)}\Psi[E]
\end{equation}
are assumed to be self-adjoint and fulfil the canonical commutation relations \eqref{eq_comm}.
In this case, there also is a mathematically rigorous version of the Hilbert space \eqref{eq_hilb_alt} and the canonical commutation relations \eqref{eq_comm} \cite{Dittrich:2012kf} arising from LQG, but the situation is more difficult since in LQG the fluxes do not commute, and hence there is no classical space $\mathcal{E}$ underlying the construction, but rather a non-commutative space. Again, in the present exploratory work, we will stay on the formal level. 

The candidate state is given by 
\begin{equation}
\Psi^{\text{LQG}}_2[E]=\int_{D(E)}  \mathcal{D}E'\,\mathcal{D}N^a\,\mathcal{D}N\: e^{-S_E[E',N^a,N]}
\end{equation}
where the integral is now over histories $E'(t,x)$ that have a common boundary value $E(x)=E'(t_0,x)$. 
These histories form a set $D(E)$. Here we assume again, as in the ADM case, that the only boundary of the manifold is $\sigma$, and that $E$ and $\dot{E}$ are regular, smooth, and give compact spacetimes (``no boundary condition''). This condition is actually very close in spirit to the original no-boundary condition of Hartle and Hawking, since $E(t,x)$ describes a 4d Euclidean geometry.

Now we can proceed as in the previous section and evaluate the action of the basic operators, and ultimately of the Hamiltonian and diffeomorphism constraints. By adding suitable boundary term, the action of $\widehat{A}$ on $\Psi^{\text{LQG}}_2$ is well defined and given by
\begin{equation}
\widehat{A}^k_a(x)\,\Psi^{\text{LQG}}_2[E]=\int_{D(E)} \mathcal{D}E\,\mathcal{D}N^a\,\mathcal{D}N\: iA_a^k(x)\, e^{-S_E[E,N^a,N]}.
\end{equation}
Then the action of the Hamiltonian constraint becomes
\begin{equation}
\widehat{H}_L\leadsto i H_E +(\alpha  A_a^mA_b^n +\beta \Gamma_a^m \Gamma_b^n
+2 \delta  A_a^m\Gamma_b^n ) 
\frac{\epsilon_{jmn} \epsilon^j{}_{kl}E^a_k E^b_l}{\sqrt{det(q)}}
\end{equation}
with 
\begin{equation}
\alpha= \frac{1+i}{\gamma^2}, \qquad \beta=(i-1)\left(1+\frac{1}{\gamma^2}\right), \qquad \text{and} \qquad \delta=(-i-1)\left(1+\frac{1}{\gamma^2}\right).
\end{equation}
The term proportional to $H_E$ integrates to zero, but the additional terms have no reason to vanish. Again there are no obvious combinations of the constraints. Therefore we conclude that $\Psi^{\text{LQG}}_2$ defined in this way \textit{does not} solve the Hamiltonian constraint. 
One can see that $\beta$ and $\delta$ vanish for $\gamma=\pm i$. This is a consistency check, as for these special values of $\gamma$ the Hamiltonian constraint simplifies considerably. Still, even for these special values, $\alpha$ remains nonzero, and the conclusion unchanged.  

We note that for the case of wave functions of $E$ considered here, there is also a problem with the diffeomorphism constraint. It will result in a term proportional to  $E^a_j A_a^mA_b^n\epsilon^j{}_{mn}$ under the integral that does not vanish. 

We note as a consistency check that also in the case of the wave functions of $E$, for a state that is of the type of the extraction amplitude, we have 
\begin{equation}
\widehat{A}(x)\leadsto   A(x), \qquad  \widehat{E}(x) \leadsto E(x), \qquad x \in \sigma  
\end{equation}
whence all the constraints are satisfied.  
\subsection{Wave function using $\Psi^{\text{HH}}_0$: $\Psi^{\text{LQG}}_3$}
Finally, we use the original HH-state in a construction of a candidate ground state for Ashtekar-Barbero variables. A plausible candidate is 
\begin{equation}
\Psi^{\text{LQG}}_3[E]:=\Psi^{\text{HH}}_0[q(E)]
\end{equation}
where we have used the fact that the 3-metric $q_{ab}$ can be expressed in terms of $E$. We see no a priori reason why this state should satisfy the constraint \eqref{eq32}, but we will check for reasons of completeness. We notice that $q(E)$ is a many-to-one function, but this is no problem. The state would just be constant along gauge orbits. Since this state is naturally a functional of $E^a_i$, we would be in the setting of the previous section, in particular $\widehat{E}$ would act as a multiplication operator, and $\widehat{A}$ as derivative, as in \eqref{eq_ops2}. 
In particular,  
\begin{equation}
\widehat{q(E)}_{ab} \leadsto q(E)_{ab}. 
\end{equation}
We need to understand how $\widehat{A}$ acts on $\Psi^{\text{LQG}}_3$. We find 
\begin{align}
\widehat{A^k_a}\,\Psi^{\text{LQG}}_3[E]&=\frac{i}{2}\frac{\delta}{\delta E^a_k}\Psi^{\text{HH}}_0[q(E)]\\
&=-\frac{1}{2i} \frac{\delta q_{bc}(E)}{\delta E_k^a} \frac{\delta}{\delta q_{bc}} \Psi^{\text{HH}}_0[q]\\
&=-\frac{1}{2}\frac{\delta q_{bc}(E)}{\delta E_k^a} \widehat{P}^{bc}\,\Psi^{\text{HH}}_0[q]\\
&=-\frac{1}{2}\frac{\delta q_{bc}(E)}{\delta E_k^a} \int_{{D}(q_{bc})}  \mathcal{D}q'\,\mathcal{D}N^d\,\mathcal{D}N\:  iP^{bc}(x)\:e^{-S_E[q',N^d,N]}\label{eq_Eaction}.
\end{align}
Using explicit expressions from \cite{Thiemann:2007zz}, we replace $P^{ab}$ with appropriate expression in terms of Ashtekar variables, which brings the right hand side of \eqref{eq_Eaction} to the form
\begin{equation}
\frac{1}{i}\frac{\delta q_{ab}(E)}{\delta E}\int\mathcal{D}[g]\,\left[|det(E^c_l)|^{-1}E^a_k E^d_k K^j_{[d} \delta^b_{c]} E^c_j\right]\,e^{-S_E[g]}.\label{eq45}
\end{equation}
In order to obtain $\delta q_{ab}(E)/\delta E$, we use the relation $q_{ab}(E)=E^j_a E^j_b |det(E^c_l)|$. However, irrespective of what $\delta q_{ab}(E)/\delta E$ would turn out to be, a look at the integrand in \eqref{eq45} indicates that we would have to replace the extrinsic curvature and rewrite it in terms of $A^i_a$ and $\Gamma^i_a$ by using \eqref{eq36}, where $\Gamma^i_a$ is again written in terms of triads and inverse triads. This replacement will turn the integrand into a really complicated expression, and we found no way to combine everything to obtain the Euclidean Hamiltonian constraint $H_E$ of LQG under the path integral. Hence we strongly suspect that the constraint is again not satisfied. 

At this point it might seem to the reader that we have simply abandoned the above line of thought as the calculations involved were complicated, and that there could have been a possibility that some of the terms involved would just cancel out, leaving behind the expected solution. However, it appears that this is not the case, as we will now explain.

The connection variables of LQG can be obtained from ADM variables by a phase space extension, followed by a canonical transformation, \cite{Ashtekar:1986yd,Barbero:1994ap,Thiemann:2007zz}.  
Instead of working directly with the LQG constraints, we could also start from scratch and try to see if the state satisfies the ADM constraints on the extended ADM phase space. If it does satisfy the constraints, we can then perform the subsequent canonical transformation. That way we could avoid the messy calculation (which we tried to justify) and arrive at a more concrete conclusion. We will sketch the calculation in the following. 

The first step is to check if the state $\Psi^{\text{HH}}_0[q(E)]$ satisfies the ADM constraints. On the extended phase space, the one-form $K^i_a$ plays the role of the canonically conjugate momentum on the spatial slice $\sigma$, provided that the Gauss constraint is satisfied. Following formal quantization, when this is promoted to a quantum operator, it takes the form
\begin{equation}
\widehat{K}^i_a=\frac{1}{i}\frac{\delta}{\delta E^a_i}=\frac{1}{i}\frac{\delta q_{bc}}{\delta E^a_i}\frac{\delta}{\delta q_{bc}}.
\end{equation}
The action of this operator on the state can be seen to be given by
\begin{align}
\widehat{K}^i_a\,\Psi^{\text{HH}}_0[q(E)]&=\frac{1}{i}\frac{\delta q_{bc}}{\delta E^a_i}\frac{\delta}{\delta q_{bc}}\,\Psi^{\text{HH}}_0[q(E)]\\
&=\frac{\delta q_{bc}}{\delta E^a_i}\,\widehat{P}^{bc}\,\Psi^{\text{HH}}_0[q]\\
&=\frac{\delta q_{bc}}{\delta E^a_i}\int_{{D}(q_{bc})}   \mathcal{D}q'\,\mathcal{D}N^d\,\mathcal{D}N\:  iP^{bc}(x)\:e^{-S_E[q',N^d,N]}\\
&=\int_{{D}(q_{bc})}  \mathcal{D}q' \,\mathcal{D}N^d \,\mathcal{D}N\,\left[\frac{i\,\delta q_{bc}}{\delta E^a_i}K^{bc}(x)\right]\,e^{-S_E[q',N^d,N]},\label{hhexps}
\end{align}
where in the last step we have assumed that the Gauss constraint holds under the path integral. This demands evaluating the term ${\delta q_{bc}}/{\delta E^a_i}$. To accomplish this, we make use of the relation between the co-triads $e^j_b$ and our electric field variable $E^b_j$. We then have
\begin{equation}
\frac{\delta q_{bc}}{\delta E^a_i}\,K^{bc}=\frac{\delta q_{bc}}{\delta e^j_d}\,\frac{\delta e^j_d}{\delta E^a_i}\,K^{bc}
= \frac{\delta e^j_d}{\delta E^a_i}\,(\delta^d_b e_{jc} + \delta^d_c e_{jb})\,K^{bc}=2 \frac{\delta e^j_d}{\delta E^a_i}\,K^d_j =2 \frac{\delta e^j_d}{\delta E^a_i}\, q^{d{d'}}\,\delta_{j{j'}}\,K^{j'}_{d'}.
\end{equation}
Using the above expression, we see that \eqref{hhexps} takes the form
\begin{equation}
\widehat{K}^i_a\,\Psi^{\text{HH}}_0[q(E)]=\int_{{D}(q_{bc})}  \mathcal{D}q' \,\mathcal{D}N^f \,\mathcal{D}N\,\left[2i \, \frac{\delta e^j_d}{\delta E^a_i}\, q^{d{d'}}\,\delta_{j{j'}} \,K^{j'}_{d'}\right]\,e^{-S_E[q',N^f,N]}.\label{hhexps1}
\end{equation}
If the action of the Lorentzian constraint on the wave function is to reproduce a multiple of the Euclidean constraint under the path integral, like in the ADM case, the action of $K$ has to lead to a term $\cancel{c}K^i_a$ for some constant $\cancel{c}$ under the path integral. 
From \eqref{hhexps1}, we see that this can only be achieved if 
\begin{equation}\label{ansatz}
2i\,\frac{\delta e^j_d}{\delta E^a_i}\, q^{d{d'}}\,\delta_{j{j'}} =\cancel{c}\,\delta^i_{j'}\,\delta^{d'}_a.
\end{equation}

While it is not too hard to calculate the functional derivative explicitly, we can use a scaling argument to show that \eqref{ansatz} does not hold. Suppose we scale the co-triads by a factor $\lambda$, that is, $e^j_d\mapsto\lambda\,e^j_d$. This means that the electric field $E^a_i$ and the inverse 3-metric scale respectively as 
\begin{equation}
E^a_i=e^a_i\,\det(e^i_a)\mapsto\lambda^{-1}\lambda^3\,e^a_i\,\det(e^i_a)=\lambda^2\,E^a_i, \qquad q_{ab}\mapsto\lambda^2\,q_{ab} \implies q^{ab}\mapsto\lambda^{-2}\,q^{ab}.
\end{equation}
This would mean that the left hand side of \eqref{ansatz} goes as $\lambda^{-3}$, whereas the right hand side stays invariant. This is a contradiction. 
Thus the state $\Psi^{\text{HH}}_0[q(E)]$ does not satisfy the ADM constraints on the extended phase space. Therefore, it would be very surprising if it satisfies the LQG constraints after the canonical transformation is performed. Hence, we do not deem it necessary to complete this procedure.

Thus we see that this observation provides a much stronger argument and we can safely conclude that the LQG constraints are not satisfied in this approach as well. 
\section{New proposal for an initial state: $\Psi^{\text{LQG}}_0$}
\label{sec_new}
In the previous section, we investigated a number of ways to translate the HH-proposal as directly as possible to Ashtekar-Barbero variables. However, all  formal states obtained in this way failed to solve the Hamiltonian constraint. Thus, to obtain a viable candidate state, it appears that we have to stay less close to the original proposal. We make the following observation: the reason the state did not satisfy the Hamiltonian constraint is because of the mismatch between the Lorentzian constraint operator, and the Euclidean constraint that is needed under the path integral. In the ADM case, the factors of $i$ picked up when the momentum acts, just effect the change to the Euclidean constraint. In Ashtekar-Barbero variables, the dependence of the constraint on the signature is more complicated, and does not coincide with a simple scaling of the momentum. This caused the extra term to pop up due to the particular placement of the signature of the manifold in the Hamiltonian constraint. If we did not want this sign change and if we had instead started with a Lorentzian action, then the formal argument of Hartle-Hawking would simply follow through and the Hamiltonian constraint (along with other constraints) would be satisfied. This observation therefore leads us to define a new initial state as
\begin{equation}\label{newp}
\Psi^{\text{LQG}}_0[A]:=\int_{D(A)}\mathcal{D}A'\,\mathcal{D}N^a\,\mathcal{D}N\:e^{-S_L[A']},
\end{equation}
where we have simply replaced Euclidean action $S_E[A]$ in \eqref{eq33} by the Lorentzian action $S_L[A]$.

The first check that needs to be done is to perform the formal calculation again that would confirm that this indeed satisfies the constraints. To this end, note that we are in the setting of section \ref{se_psiofa}.  We thus have
\begin{equation}
\widehat{A}\leadsto A, \qquad \widehat{E} \leadsto iE.  
\end{equation}
Furthermore, we had already seen in section \ref{se_psiofa} that 
\begin{equation}
\widehat{K}\leadsto K. 
\end{equation}
Putting everything together, we obtain
\begin{equation}
\widehat{H}_L \equiv [\widehat{F}^j_{ab}\,-\,({\gamma}^2 +1)\epsilon_{jmn}\,\widehat{K}_a^m \widehat{K}_b^n]\frac{\epsilon_{jkl}\widehat{E}^a_k \widehat{E}^b_l}{\sqrt{det(\widehat{q})}} \leadsto -\sqrt{i}\,H_L .
\end{equation}
Thus we see that the modified Hartle-Hawking state in \eqref{newp} vanishes under the path integral, implying that the state \textit{does satisfy} the Lorentzian Hamiltonian constraint. The same holds for Gauss and diffeomorphism constraints, since 
\begin{equation}\label{eq_gaussdiff}
\widehat{G}_j\leadsto i G_j, \qquad \widehat{H}_a\leadsto i H_a.
\end{equation}
Then the action of the constraints can be rewritten as a derivative with respect to lapse function, shift vector, and the Lagrange multiplier with respect to the Gauss constraint as in \eqref{eq_a}. Translation invariance of the respective functional integral measures then shows that the action of all the constraints is zero.  
Thus the new state fulfills the minimum requirements. 

One immediate question is regarding the precise definition of the domain $D(A)$. In this case the histories describe, albeit in an indirect way, a Lorentzian geometry. Therefore there can not be compact spacetime histories that have no boundaries beyond $\sigma$ and are regular. We will not decide on a replacement of the no-boundary condition at this point, but note that it makes the present proposal somewhat incomplete and less natural than the original one. 

To get a feeling for the physics implied by the new state, we will consider it in the context of quantum cosmology in the following section. 
\section{The new state in quantum cosmology}
\label{sec_lqc}
The Hartle-Hawking state is a formal state for the full theory of quantum gravity. Since it is formally defined by a complicated path integral, it is hard to use it for any kind of concrete calculation. To simplify things, one can reduce the number of degrees of freedom, by just considering a symmetric sector of GR with finitely many degrees of freedom and quantizing the corresponding phase space. Then the HH prescription yields a state that is defined by a quantum mechanical path integral, which is under relatively good control, and detailed calculations become possible.   
This was done by Hartle, Hawking and others for the case of ADM variables. One has to note that here, as in other cases, it is by no means clear that quantization and symmetry reduction commute, even in some approximate sense. That is, a HH state for a quantization of the cosmological sector of GR is not necessarily close to the state one obtains by restricting the HH state \eqref{eq9} of the full theory to symmetric configurations $q_{ab}$. 

With this proviso, we will now consider cosmology in Ashtekar-Barbero-like variables. This is the starting point of Loop Quantum Cosmology (LQC) \cite{Bojowald:2001xe,Ashtekar:2003hd,Ashtekar:2011ni}, which is obtained by applying the principles of LQG to cosmological settings. From here on, we set $8\pi G=1$, and use the notation and results from \cite{Ashtekar:2011ni} in the following.

We introduce a cubical fiducial cell $\mathcal{C}$ as we would like to equip it with a fiducial metric $\mathring{q}_{ab}$ of Euclidean signature. This comes as a requirement as we are mimicing the quantization procedure of a background independent theory. Moreover, let $\mathring{e}^a_i$ and $\mathring{\omega}^i_a$ be the associated orthonormal frames and co-frames. The symmetries of Friedmann-Lema\^{\i}tre-Robertson-Walker (FLRW) spacetimes imply that from each equivalence class of gauge related homogeneous, isotropic pairs $(A^i_a,E^a_i)$, we can select one, such that \cite{Ashtekar:2003hd}
\begin{equation}
A^i_a=\tilde{c}\,\mathring{\omega}^i_a \qquad\text{and}\qquad E^a_i=\tilde{p}(\mathring{q})^{\frac{1}{2}}\,\mathring{e}^a_i.
\end{equation}

This means that the information contained in the above canonical pair is now captured by just two functions of time, namely $\tilde{c}$ and $\tilde{p}$. In terms of scale factor $a$ and $\gamma>0$ being the Immirzi parameter of LQG, they are given by $\tilde{c}=\gamma\dot{a}$ and $\tilde{p}=a^2$. In LQG, due to the gauge freedom that exists in the triads, the ADM phase space is enlarged, and yet keeps all the dynamics of GR intact. This is expected to turn up when considering FLRW models as well. In other words, in addition to the property of being degenerate, its possible for the physical triads to have both orientations. On this full space, with $\tilde{p}\in\mathbb{R}$, we have three possibilities: $\tilde{p}>0$ if $E^a_i$ and $\mathring{e}^a_i$ have the same orientation, $\tilde{p}<0$ if the orientations are opposite, and $\tilde{p}p=0$ is $E^a_i$ is degenerate. 

Therefore, the LQC phase space is coordinatized by the quadruplet $(\tilde{c},\tilde{p};\phi,p_\phi)$ with non-zero Poisson brackets given by
\begin{equation}\label{sf}
\{\tilde{c},\,\tilde{p}\}=\frac{\gamma}{3V_o} \qquad\text{and}\qquad \{\phi,p_\phi\}=1,
\end{equation}
where $V_o$ is the volume of the fiducial cell $\mathcal{C}$ with respect to the fiducial metric $\mathring{q}_{ab}$. Note here that the phase space and therefore the symplectic structure carries a cell dependence. Following \cite{Ashtekar:2003hd}, it is mathematically convenient to rescale the canonical variables as follows: Set $c:=V_o^{\frac{1}{2}}\tilde{c}$ and $p:=V_o^{\frac{2}{3}}\tilde{p}$, so that
\begin{equation}\label{cpcr}
 \{c,\,p\}=\frac{\gamma}{3}. 
\end{equation}
Then $c, p$ are insensitive to the choice of $\mathring{q}_{ab}$ and the Poisson bracket between them. Due to the underlying symmetries and the gauge fixing, only the Hamiltonian constraint remains which is now given by:
\begin{equation}\label{eq_miniham}
H_L=|p|^{\frac{1}{2}}\left(\frac{p_\phi^2}{2p^2} - \frac{3}{\gamma^2}c^2+\frac{\Lambda}{2}|p|\right)\approx 0.
\end{equation} 
The Lorentzian action can be written as 
\begin{equation}
\label{eq_miniaction}
S^{(c,p)}_L[N,c,\phi]=\int \text{d}t\;\left( \frac{3}{\gamma}\,p\,\dot c(t)+p_\phi\,\dot{\phi}(t) - N(t) H_L(t)  \right)
\end{equation}
where we understand $\phi(t),p_\phi(t)$ as functions of $c,p$ by virtue of the equations of motion. The gravitational variables $c, p$ are directly related to the basic canonical pair $(A^i_a, E^a_i)$ in
full LQG and enable one to introduce a quantization procedure in LQC that closely mimics LQG. 
We will not use this quantization here, but stay completely formal. The Hilbert space we choose is
\begin{equation}
\mathcal{H}=L^2(\mathbb{R}^2,\text{d}c\,\text{d}\phi).
\end{equation}
Basic operators are 
\begin{equation}
\widehat{c}=c, \qquad \widehat{p}=\frac{1}{i}\frac{\gamma}{3} \frac{\partial}{\partial c}, \qquad   
\widehat{\phi}=\phi, \qquad \widehat{p_\phi}=\frac{1}{i} \frac{\partial}{\partial \phi}. 
\end{equation}
To stay as closely as possible to the definition \eqref{newp}, we define the state as a wave function of $c$, i.e.,  
\begin{equation}
\Psi^{(c,p)}_0[c,\phi]:= \int_{D(c,\phi)} \mathcal{D}c'\,\mathcal{D}\phi'\,\mathcal{D}N\; e^{-S^{(c,p)}_L[c',\phi']}
\end{equation}
where the superscript refers to the fact that we are dealing with an LQC state based on the canonical action for $c$ and $p$. 

The quantum dynamics of the FLRW model is significantly simplified in terms of a slightly different pair of canonically conjugate variables, $(b, v)$, for the gravitational field.
These variables are given by 
\begin{equation}\label{gv}
b:=\frac{c}{|p|^{\frac{1}{2}}},\qquad v:=4|p|^{\frac{3}{2}}\,\text{sgn}(p)\qquad\text{so that}\qquad \{b,\,v\}=2\gamma
\end{equation}
where $\text{sgn}(p)$ is the sign of $p$ ($1$ if the physical triad $e^a_i$ has the same orientation as the fiducial $\mathring{e}^a_i$ and $-1$ if the orientation is opposite). In terms of this pair, the Hamiltonian constraint takes the form:
\begin{equation}\label{hc}
H_L=|v|\left(\frac{2p_\phi^2}{v^2}-\frac{3}{4\gamma^2}b^2+\frac{\Lambda}{8}\right)\approx 0
\end{equation}
and we obtain a canonical action 
\begin{equation}
S_L^{(b,v)}=\int_i^o dt \left(p_{\phi}\dot{\phi} + \frac{v\dot{b}}{2\gamma} - H_L^{(b,v)}\right).
\end{equation}
While the variables $b,v$ are not as closely related to the variables of the full theory, they are widely used and it is thus of interest to consider the state 
\begin{equation}
\Psi^{(b,v)}_0[b,\phi]:= \int_{D(b,\phi)} \mathcal{D}b'\,\mathcal{D}\phi'\,\mathcal{D}N\; e^{-S^{(b,v)}_L[b',\phi']}.
\end{equation}
As we will see more explicitly below, this action and \eqref{eq_miniaction} differ by a boundary term, the generating function for the canonical transformation $(c,p) \rightarrow (b,v)$. Therefore the two states these variables define are genuinely different. 

The first check that we perform is to see if the proposed states satisfies the quantum Lorentzian constraint. Using \eqref{eq_miniaction} and the arguments presented in the full theory, we have for $\Psi_0^{(c,p)}$:
\begin{equation}
\widehat{c}\leadsto c, \qquad \widehat{p}\leadsto ip, \qquad
\widehat{\phi}\leadsto \phi, \qquad \widehat{p_\phi}\leadsto ip_\phi\,.
\end{equation}
From the form \eqref{eq_miniham} we see that 
\begin{equation}
\widehat{H_L}\leadsto H_L
\end{equation}
whence the state indeed satisfies the Lorentzian Hamiltonian constraint as in the full theory. 

The arguments for $\Psi^{(b,v)}_0$ are completely analogous, with the Hilbert space and operators for $b$ and $v$ defined in the same way as those for $c$ and $p$. Again the conclusion is that the state $\Psi_0^{(b,v)}$ satisfies the constraints in a formal sense. 

To get a handle on possible physical implications of the states, in what follows, we perform a saddle point approximation and study cases with zero cosmological constant and with a positive cosmological constant. The gist of the method is that the largest contribution to the path integral comes from the stationary points of the action. In our case, the formal result would be 
\begin{equation}
\label{eq_saddle}
\Psi^{(c,p)}_0(c_o,\phi_o)\approx \frac{1}{\sqrt{|\det (S''_L[X]/2\pi)|}} e^{-S^{(c,p)}_L[X]}
\end{equation}
where $X$ is a critical point of $S^{(c,p)}_L$, 
\begin{equation}
\label{eq_crit}
\delta S_L^{(c,p)}[X]=0 
\end{equation}
and an analogous approximation for $\Psi^{(b,v)}_0$. 

$S''_L|_X$ denotes the Hessian of $S_L$ at the critical point. $X$ may be a complex critical point, but one has to assume that it is a minimum of the real part of $S_L$. In the case of several critical points, the integral would be approximated by a sum of terms of the same form as the right hand side of \eqref{eq_saddle}. The approximation is expected to be leading order in $\hbar$, becoming better with $\hbar\rightarrow 0$. 

In our case, $X$ is subject to boundary conditions. Since we are in the Lorentzian domain, there are two boundaries, which we will denote by $i$ (initial) and $o$ (outgoing) in the following. One set of conditions is that $X$ has the arguments $c_o,\phi_o$, or $b_o,\phi_o$, of the wave function as boundary values at the outgoing slice. Since \eqref{eq_crit} are the equations of motion, we expect $X$ to depend on two more parameters $x,y$, so
\begin{equation}
X\equiv X(c_o,\phi_o,x,y),\quad \text{ or }\quad X\equiv X(b_o,\phi_o,x,y). 
\end{equation}
We could chose $x,y$ to be further boundary values at the slices $i$ or $o$.  

The statement  \eqref{eq_saddle} is completely formal because one would have to assume some form of functional analyticity of $S$, and there is a priori no definition of $\text{det}(S''_L/2\pi)$ that makes sense, since $S''_L$ is infinite dimensional. As a consequence, and in the spirit of the entire article, we will be very coarse about the calculation. In particular, we will not check that the critical points we find are minima of the real part, and we will completely drop $\text{det}(S''_L/2\pi)$. Also, at least if written as above, the actions $S^{(c,p)}_L$, $S^{(b,v)}_L$ are not analytic in an obvious sense.  



In the following, since the result for non-zero $\Lambda$ seems to be continuous for $\Lambda\rightarrow 0$, we will treat both $\Lambda=0$ and $\Lambda > 0$ case in a unified way in terms of each canonical pair. As indicated before, we will see that these two pairs differ by a boundary contribution in their respective actions. 
\subsection{Saddle point approximation for the canonical pair $(c,p)$ }
From \eqref{eq_miniham}, we have the Hamiltonian constraint written in terms of the variables $(c,p)$ as
\begin{equation}
H_L^{(c,p)}=\left(\frac{p_\phi^2}{2|p|^{3/2}} - \frac{3}{\gamma^2}c^2\sqrt{|p|}+\frac{\Lambda}{2}|p|^{3/2}\right)\approx 0.
\end{equation} 
Using the commutation relations of the scalar field \eqref{sf} and that of gravitational variables \eqref{cpcr}, we compute the expressions for $\dot{\phi}$ and $\dot{c}$ as
\begin{equation}\label{dots1}
\dot{\phi}=\frac{\partial H_L^{(c,p)}}{\partial p_{\phi}}=\frac{p_{\phi}}{|p|^{3/2}}\qquad\text{and}\qquad \dot{c}=\frac{\gamma}{3}\,\frac{\partial H_L^{(c,p)}}{\partial p}=\frac{\gamma}{3}\left(-\frac{3p_{\phi}^2}{4|p|^{5/2}}-\frac{3c^2}{2\gamma^2\sqrt{|p|}}+\frac{3\Lambda\sqrt{|p|}}{4}\right).
\end{equation}
We have the canonical action as
\begin{equation}
S^{(c,p)}=\int_i^o dt \left(p_{\phi}\dot{\phi} + \frac{3}{\gamma}p\dot{c} - H_L^{(c,p)}\right).
\end{equation}
On-shell, i.e., on the constraint surface, the classical Hamiltonian constraint becomes an equality and therefore, using \eqref{dots1}, the action $S$ can be evaluated as
\begin{align}
\label{actcp} S^{(c,p)}&=\int_i^o dt \left(p_{\phi}\dot{\phi} + \frac{3}{\gamma}p\dot{c}\right)\\ 
&=\int dt \left(\frac{p_{\phi}^2}{|p|^{3/2}}- \frac{3c^2\sqrt{|p|}}{2\gamma^2}+\frac{3\Lambda}{4}|p|^{3/2}\right)\\
&=\int dt \left(\frac{1}{2}H_L^{(c,p)} + \frac{\Lambda}{2}|p|^{3/2}\right)\\
&=\frac{\Lambda}{2}\int dt\;|p|^{3/2}.\label{s1}
\end{align}

To evaluate this term explicitly, we do integration by parts of the action \eqref{actcp}. A similar procedure as above leads to
\begin{equation}
S^{(c,p)}=-\Lambda \int_i^o dt \; |p|^{3/2} + \frac{3}{\gamma}\left(c_o p_o - c_i p_i\right).\label{s2}
\end{equation}
Thus, equating \eqref{s1} and \eqref{s2} results in the simplification of the bulk term of the action which turns out to be proportional to the boundary term:
\begin{equation}
S^{(c,p)}=\frac{\Lambda}{2}\int_i^o dt\; |p|^{3/2} = \frac{1}{\gamma}\left(c_o p_o - c_i p_i\right).
\end{equation}

Notice that for the vanishing cosmological constant scenario, the bulk term in \eqref{s1} vanishes, which is equivalent to $c_o p_o=c_i p_i$. This equality can be seen to arise from the Hamiltonian constraint since (on the constraint surface) the product $pc$ on each boundary is proportional to $p_{\phi}$, which is a constant. 

\subsection{Saddle point approximation for the canonical pair $(b,v)$}
From \eqref{hc}, we have the Hamiltonian constraint written in terms of the variables $(b,v)$ as
\begin{equation}
H_L^{(b,v)}=\left(\frac{2p_\phi^2}{|v|}-\frac{3}{4\gamma^2}b^2|v|+\frac{\Lambda |v|}{8}\right)\approx 0.
\end{equation}
Using the commutation relations of the scalar field \eqref{sf} and that of gravitational variables \eqref{gv}, we compute the expressions for $\dot{\phi}$ and $\dot{b}$ as
\begin{equation}\label{dots2}
	\dot{\phi}=\frac{\partial H_L^{(b,v)}}{\partial p_{\phi}}=\frac{4p_{\phi}}{v}\qquad\text{and}\qquad \dot{b}=2\gamma\frac{\partial H_L^{(b,v)}}{\partial v}=2\gamma\left(-\frac{2p_{\phi}^2}{v^2}-\frac{3b^2}{4\gamma^2}+\frac{\Lambda}{8}\right).
\end{equation}
In this case, our canonical action is
\begin{equation}
S^{(b,v)}=\int_i^o dt \left(p_{\phi}\dot{\phi} + \frac{v\dot{b}}{2\gamma} - H_L^{(b,v)}\right).
\end{equation}
As before, on the constraint surface, the classical Hamiltonian constraint becomes an equality and therefore, using \eqref{dots2}, the action $S$ can be evaluated as
\begin{align}
S^{(b,v)}&=\int dt \left(p_{\phi}\dot{\phi} + \frac{v\dot{b}}{2\gamma}\right)\\
&=\int dt \left(\frac{4p_{\phi}^2}{v}-\frac{2p_{\phi}^2}{v}-\frac{3b^2 v}{4\gamma^2}+\frac{\Lambda v}{8}\right)\\
&=\int dt \left(\frac{2p_{\phi}^2}{v}-\frac{3b^2 v}{4\gamma^2}+\frac{\Lambda v}{8}\right)\\
&=\int dt\;(H_L)\:,
\end{align}
which vanishes on the constraint surface. This implies that the action $S^{(b,v)}=0$ and therefore the state is flat in connection representation.

Notice that this action remains zero even in the case when $\Lambda=0$. Therefore, we conclude that the quantum state for LQC models in $(b,v)$ variables is quite reminiscent of the Ashtekar-Lewandowski vacuum, which is a flat functional of the connection.

\subsection{Comparison of LQC models in terms of the two canonical pairs}
In the two sections above, we have evaluated the canonical action in terms of both pairs of variables that are used for quantization in LQC models. 

An important observation is that even though the two pairs of variables are completely equivalent to each other, the states they generate according to our adaptation of the Hartle-Hawking prescription are in general not. The actions differ by a boundary term, the generating function of the canonical transformation. 
In the setting that we have considered, the spatial curvature vanishes, and so do the bulk contributions. Moreover, the boundary term also vanishes in the case $\Lambda =0$. In that case, the states have the same functional form. Moreover, the action $S^{(c,p)}$ can indeed be equivalently written in terms of $(b,v)$ variables as
\begin{equation}
S^{(c,p)}=\frac{1}{\gamma}(c_o p_o - c_i p_i)=\frac{1}{\gamma}(b_o v_o - b_i v_i).
\end{equation}
This will be helpful in analyzing $\Psi_0^{(c,p)}$ further. 

Of all the cases we have considered, only $\Psi^{(c,p)}_0$ fo $\Lambda\neq 0$ has a non-trivial form in the saddle point approximation. With an appropriate choice of a boundary condition for $c_i$, its possible to re-write $p_o$ in terms of $c_o$ and $p_{\phi}$ using the constraint equation. This can be the starting-point of an investigation of further properties of $\Psi^{(c,p)}_0$. We leave this investigation for another time. 


\section{Conclusions}
In this work, we have studied the question of how the formal ground state for canonical quantum gravity proposed by Hartle and Hawking could be applied in loop quantum gravity. The most important result of the work seems to be the fact, that an immediate translation of the construction of Hartle and Hawking to the variables used in loop quantum gravity is not possible, since the resulting state does not have the same formal properties as the original. In particular, the original state is constructed from the \emph{Euclidean} action, but still satisfies the \emph{Lorentzian} Hamiltonian constraint. However, this is no longer the case with the analogous construction of the state in terms of Ashtekar-Barbero variables, implying that the construction seems to be dependent on the choice of variables used in the quantization. This seems to hold true even for analogous but much simpler systems such as the relativistic particle and thus makes us confident in our conclusion. The fact that the construction works for ADM variables but not for some others can be interpreted as nature taking a preference in them. One can also take this fact as a mere coincidence without further ramifications. In any case, it is an interesting observation. 

With the obvious simplest generalization off the table, we have looked for alternatives that do satisfy all the quantum constraints at least in a formal sense. A possibility we found is to use the Lorentzian action in place of the Euclidean one in the construction of the state. The resulting state satisfies all the quantum constraints in a formal sense. However, with integrating over Lorentzian spacetimes, one needs to talk about boundary conditions again, since these geometries can not be compact anymore. Thus some of the elegance of the ``no-boundary'' proposal is lost.

We have investigated the new proposal in some detail in the cosmological setting. We looked at spatially flat FLRW cosmology with and without positive cosmological constant, and for two sets of canonical variables. The different variables yield, in general, different states. This underscores the observation made in the full theory that the construction of Hartle-Hawking-like states is dependent on the choice of the canonical variables. 

In a formal saddle point approximation, we found that for the special case of $\Lambda=0$ the states coincide, and are independent of $c$, $b$, respectively. This is very reminiscent of the state that is a ground state in loop quantum gravity, which is completely flat in $A$. For $\Lambda>0$, the two states diverge from each other, with $\Psi_0^{(b,v)}$ staying flat in the saddle point approximation, and with $\Psi_0^{(c,p)}$ developing a non-trivial $c$-dependence.
 

There are several loose ends. On the one hand, the state $\Psi_0^{(c,p)}$ for $\Lambda>0$ should be investigated more carefully.  One important outstanding check is to demonstrate that the result of the saddle point approximation indeed satisfies the quantum constraints in a suitable approximate sense. On the other hand, one could ponder the failure of the ``obvious'' generalization of the Hartle-Hawking construction to Ashtekar-Barbero variables more deeply. Signature change is more complicated in the corresponding Hamilton constraint, in particular it is intertwined with the Immirzi parameter. It would thus be interesting to consider other generalizations, possibly involving changes in the Immirzi parameter, that satisfy the Lorentzian constraint through a more complicated mechanism than our present proposal. 

It is an intriguing observation that our generalization of the Hartle-Hawking state approximately reproduces the Ashtekar-Lewandowski vacuum in certain cases. Whether this is an accident of the cosmological models that we considered or whether it has a deeper meaning remains to be seen as well.  

\begin{acknowledgments}
SD and HS thank their colleagues at the Institute for Quantum Gravity, FAU Erlangen-N\"urnberg for discussions. The question of how to apply the Hartle-Hawking proposal to LQG was first brought up by D.~Yeom, and HS thanks him and D.~Hwang for many enlightening discussions about the Hartle-Hawking state. 
\end{acknowledgments}

\appendix*
\section{Free Relativistic Particle -- a toy model}
\label{sec_particle}
The canonical formulation of ``Polyakov action'' for the free relativistic particle shares some similarities with general relativity in Ashtekar-Barbero variables. It is a reparametrization invariant theory, leading to a constrained canonical description. Moreover, as in Ashtekar-Barbero variables, two metrics (world line metric, target space metric) play a role in the theory, and we can express the world line metric by an einbein field.  
Therefore, we will consider the free relativistic particle as a toy model in this appendix. We will show that an analogue of the HH state runs into problems even in this very simple setting. We will perform the same procedure we have followed for the ADM and Ashtekar variables. 

For a free relativistic particle, the action is given by 
\begin{equation}
\int d\tau\:m\:\sqrt{-\eta_{\mu\nu}\dot{x}^\mu\,\dot{x}^\nu}
\end{equation}
where $\eta_{\mu\nu}$ is the Minkowski metric and $m$ is the mass of the particle. 

In analogy with Palatini/tetrad action, consider the following action with $e(\tau)$ as the (absolute value of the) einbein field along the world-line of the particle \cite{Green:1987sp}:
\begin{equation}\label{eq47}
S_L=\frac{1}{2}\int d\tau\,(e^{-1}\eta_{\mu\nu}\dot{x}^\mu\,\dot{x}^\nu\,-em^2).
\end{equation}
It can easily be checked that this action is invariant under reparametrizations. For this action, the conjugate momenta are given by:
\begin{align}
P_\mu &= \frac{\dot{x}^\nu\eta_{\mu\nu}}{e}\equiv \frac{\dot{x}_\nu}{e} ,\\
P_e &= 0.
\end{align}
Setting $\dot{e}=v^e$, the Hamiltonian becomes
\begin{equation}
H_L = (P_{\mu}\dot{x}^\mu\,+P_e v^e\,-L)=\frac{e}{2}(P_\mu P^\mu\,+ m^2),
\end{equation}
where $L$ is the Lagrangian. $e$ plays the role of a Lagrange multiplier, enforcing the constraint
\begin{equation}\label{eq52}
H \equiv \frac{e}{2}(P_\mu P_\nu \eta^{\mu\nu}  \,+ m^2)=0.
\end{equation}
A similar calculation can be carried out for the Euclidean theory, which we take to be defined by the Wick rotation 
$t \rightarrow -i\tau$ and $\eta_{\mu\nu} \rightarrow \delta_{\mu\nu}$, i.e.,
\begin{equation}
S_E = \frac{1}{2} \int d\tau\:(e^{-1}(\tau)\delta_{\mu\nu}\dot{x}^\mu\,\dot{x}^\nu\,+e(\tau)m^2).
\end{equation}
The canonical momenta are now 
\begin{align}
P_\mu &= \frac{\dot{x}^\nu\delta_{\mu\nu}}{e},\\
P_e &= 0,
\end{align}
which we have not distinguished by notation from their Lorentzian counterparts. In principle, one has to carefully work with the different dependence of the momenta on the velocities for the different signatures in the following calculation, but it turns out that due to the quadratic nature of the action, this subtlety does not have any effect. In the Euclidean theory, the constraint is 
\begin{equation}
H_E=  \frac{e}{2}(P_\mu P_\nu \delta^{\mu\nu} \,+ m^2)=0. 
\end{equation}
When splitting into components, we thus have 
\begin{align}
H_L&= -\frac{e}{2}(P_0)^2\,+\frac{e}{2}(P_i)^2\,+\frac{em^2}{2},\\
H_E&= \frac{e}{2}(P_0)^2\,+\frac{e}{2}(P_i)^2\,-\frac{em^2}{2}.
\end{align}
Formal quantization has 
\begin{equation}
\widehat{P}_0= \frac{1}{i}\frac{\delta}{\delta {x}^0},\qquad 
\widehat{P}_k= \frac{1}{i}\frac{\delta}{\delta {x}^k}.
\end{equation}
By suitably choosing the discretization of the path integral at the boundary, or equivalently by adding a suitable boundary term, we obtain 
\begin{equation}
\widehat{P}_\mu \leadsto iP_\mu\equiv \frac{i \dot{x}^\nu\delta_{\mu\nu}}{e}. 
\end{equation}
For 
\begin{equation}
\Psi^{\text{particle}}_0[x]:= \int_{D(x)}\mathcal{D}x'\:\mathcal{D}e\:e^{-S_E[x',e]},
\end{equation}
we then find 
\begin{align}
\widehat{H}_L\,\Psi^{\text{particle}}_0[x]&=\left[-\frac{e}{2}(\widehat{P}_0)^2\,+\frac{e}{2}(\widehat{P}_i)^2\,+\frac{e\,m^2}{2}\right]\: \int_{D(x)}\mathcal{D}x'\:\mathcal{D}e\:e^{-S_E[x',e]}\\
&= \int_{D(x)}\mathcal{D}x'\:\mathcal{D}e\:\left[\frac{e}{2}(P_0)^2\,-\frac{e}{2}(P_i)^2\,+\frac{e\,m^2}{2}\right]\:e^{-S_E[x',e]}\\
&= \int_{D(x)}\mathcal{D}x'\:\mathcal{D}e\:\left[-H_E\,+ e\,(P_0)^2\right]\: e^{-S_E[x',e]}.
\end{align}
Thus, apart from the vanishing Euclidean Hamiltonian constraint, we obtain an extra term which, however, does not go to zero under the path integral. Comparing the relativistic particle with the LQG case, we notice that these two cases are quite similar, in the sense that in both these cases there appears an extra term which does not vanish under the path integral.


\end{document}